\documentclass[preprint2]{aastex}  

\def\Msun{\mbox{M$_\odot$}}

\def\mathnew{\mathsurround=0pt}   
\def\simov#1#2{\lower .5pt\vbox{\baselineskip0pt  
    \lineskip-.5pt\ialign{$\mathnew#1\hfil##\hfil$\crcr#2\crcr\sim\crcr}}}

\def\'#1{\ifx#1i{\accent"13\i}\else{\accent"13#1}\fi}

\begin{document}    
\shorttitle{Mass Distribution in NGC2976}     
\shortauthors{Valenzuela et al. 2013}

\title{Non-Axisymmetric  Structure in the Satellite  Dwarf Galaxy NGC2976: Implications for its Dark/Bright  Mass Distribution and Evolution}             

\author{Octavio Valenzuela$^{1}$, Hector Hernandez-Toledo$^{1}$, Mariana Cano-D\'iaz$^{1,2}$, Ivanio Puerari$^{3}$, Ronald Buta$^{4}$, B\'arbara Pichardo$^{1}$, Robert  Groess$^{5}$ }

\affil{$^1$ Instituto de Astronom\'ia,Universidad Nacional Auton\'oma de Mexico, A.P. 70-264,
04510, M\'exico, D.F.,octavio@astro.unam.mx\\ $^2$  Osservatorio Astronomico di Roma - INAF. Via di Frascati 33, 00040 Monte Porzio Catone, Italy\\ $^3$  Instituto Nacional de Astrof\'isica Optica y Electr\'onica. Calle Luis Enrique Erro 1, 72840, Sta. Maria Tonantzintla, Puebla, Mexico \\$^4$ Department of Physics and Astronomy, University of Alabama, Tuscaloosa, AL 35487, USA. \\$^5$ School of Computational and Applied Mathematics, University of Witwatersrand, Private Bag 3, WITS 2050, South Africa }  

\begin{abstract}  
We present the result of an extensive search for non-axisymmetric structures in the dwarf satellite galaxy of M81: NGC 2976, using multiwavelength archival observations. The galaxy is known to present   kinematic evidence for a bysimmetric distortion \citep{Spekkens07,SellwoodZam2010, Adams2976}  however  the stellar bar presence is controversial. This controversy  motivated the possible interpretation of NGC 2976 presenting an elliptical disk triggered by a prolate dark matter halo \citep{Simon03, KazantzidisTriax}.    We applied  diagnostics  used in  spiral galaxies  in order to detect stellar bars or spiral arms.  The m=2 fourier phase has a jump around 60  arcsecs consistent with a central bar and bisymmetric arms. The CO, 3.6 $\mu$ surface brightness and the dust lanes are consistent with a gas rich central  bar and possibly with gaseous spiral arms.  The barlike feature is close to 20 degrees offset the disc position angle, in agreement with kinematic estimations. The  kinematic jumps related with the dust lanes  suggest that  the bar perturbation in the disk kinematics is non-negligible and the reported non-circular motions, the  central gas excess and the  nuclear X-ray source (AGN/Starburst) might be produced by the central bar.  SPH  simulations of disks inside triaxial dark halos suggest that the two  symmetric spots at  130 arcsecs and the narrow arms may be produced by gas at turning points in an elliptical disk, alternatively the potential ellipticity can be originated by  tidally induced strong stellar bar/arms, in both cases rotation curve interpretation is importantly biased.   The M81 group is a natural candidate to trigger the bisymmetric distortion and the related evolution as suggested by the  HI tidal bridge detected by  \citet{Chynoweth2008}.  We conclude that both mechanisms,   the gas rich bar and spiral arms  triggered by environment (tidal stirring), and  primordial halo triaxiality,  can explain most of NGC 2976 non-circular motions,  mass redistribution and nuclear activity. Distinguishing between them requires detailed modeling of environmental effects.  A similar analysis like ours may reveal such kind of structures in other nearby dwarf satellite galaxies, if this is confirmed, the same evolutionary scenario will be applicable to them. This implies biases constraining their dark matter distribution and also making comparison against theoretical predictions for isolated galaxies. 
\end{abstract}                
 
\section{Introduction}                                                     
\label{sec:intro}

The internal kinematics of galaxies has been one of the central elements that lead to initially establish  the dark matter hypothesis in astrophysics \citep{Rubin1978, Bosma1978PhD}.  In particular the non-keplerian behavior  of  disk  galaxies rotation  curves  has been naturally incorporated into the current  cosmological paradigm, the LCDM model, through  dark matter halos surrounding galaxies, and even the halo mass assembly  is accurately  predicted by the model \citep[e.g.]{Frenk88}.  The assumption of a nearly full rotational support of disk galaxies is critical in order to compare cosmological predictions with data.  Taking this  comparison strategy further in detail it has been found that LSB/dwarf galaxies may reveal a potential flaw for the most simple cosmological predictions for rotation curve shapes that neglect baryonic physics, the so called core-cusp problem  \citep{Moore1994, RFlores1994, dBlokAdAst2010} . Including the effect of baryons although challenging, offers possible solutions like modifying the internal halo structure  \citep{Mashchenko2006, Governato2010N}  or biasing the interpretation of kinematics because of galaxy structure or pressure support \citep{Rhee04, valenzuela07}.  The possibility of distinguishing between different  galaxy formation scenarios makes important and interesting to quantitatively evaluate  possible deviations from the rotational support assumption, the so called  non-circular motions.  They have been largely discussed either motivated by the observed structure of some galaxies like bars, spirals, lopsidedness \citep{Rhee04, Spekkens07,SellwoodZam2010} or  the possibility of elliptical galaxy disks triggered by triaxial dark matter halos but masked by projection effects \citep{Hayashi06}.    The pursue for adopting methods free from the non-circular motions complications  has   triggered a revision of both theoretical \citep{Bailin07}  and observational   analysis techniques \citep{Kuzio2012, Kuzio2011, Spekkens07}.  Currently in observations, the presence of non-circular motions are a common result, besides almost every new study reports galaxies whose rotation curves are consistent with cores but also a minority consistent with cusps \citep{Simon05}, in some cases cuspier than cosmological predictions,  suggesting some systematic effect.  However, still  there is not an unanimous agreement about  the interpretation of these observational results, in particular for the detected non-circular motions, for example some valuable methods assume the epicyclic approximation which may be problematic for strong perturbation like bars or considerably elliptical disks \citep{Franx1994, Schoenmakers1997}, others do not adopt these hipothesis \citep{Spekkens07, SellwoodZam2010, Kuzio2012}. Some other methods assume a subdominant and random nature to non-circular motions  (Oh et al, 2011).   
  
\begin{figure} [h!] 
\epsscale{0.75}
\plotone{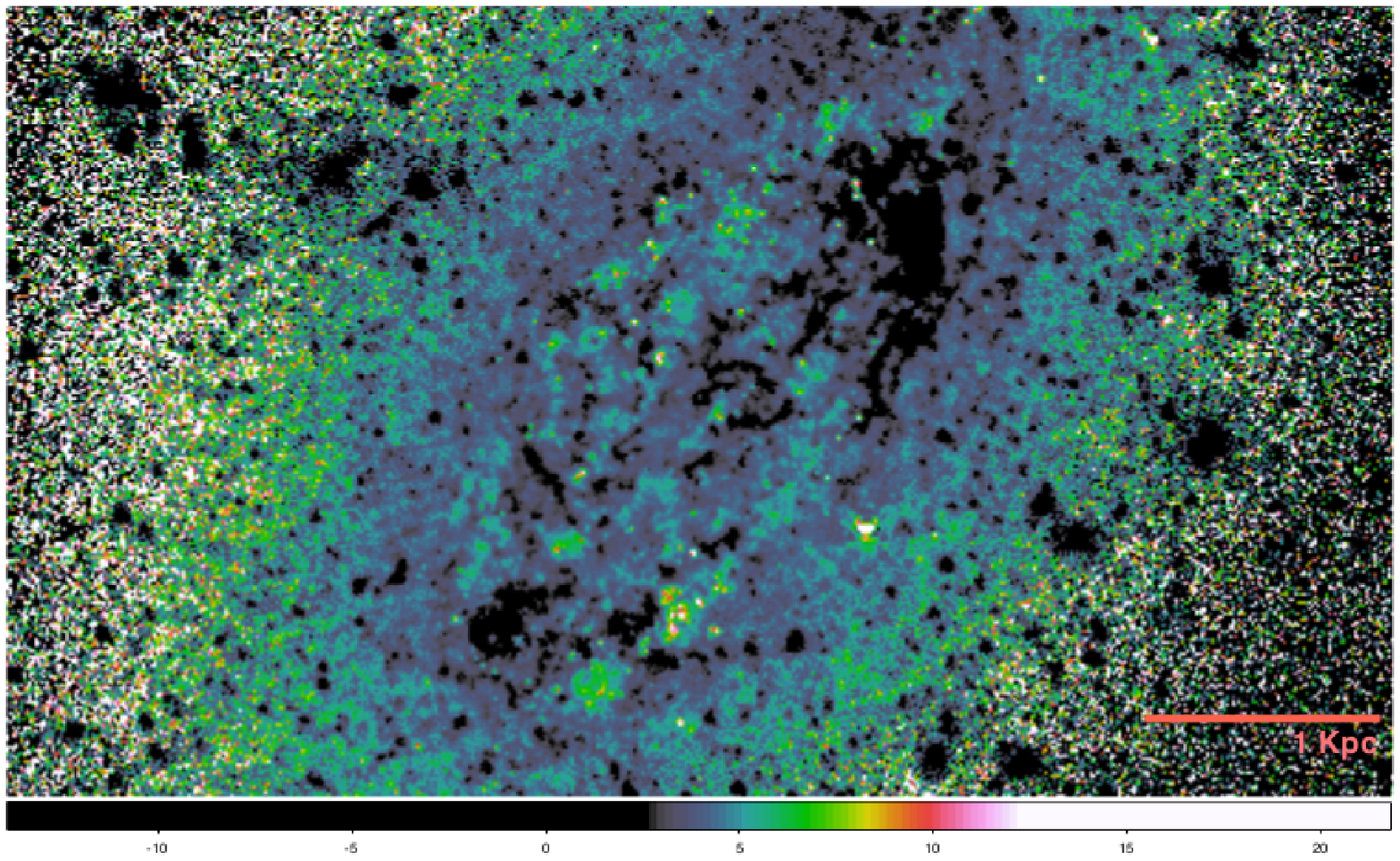}
\epsscale{0.75}
\plotone{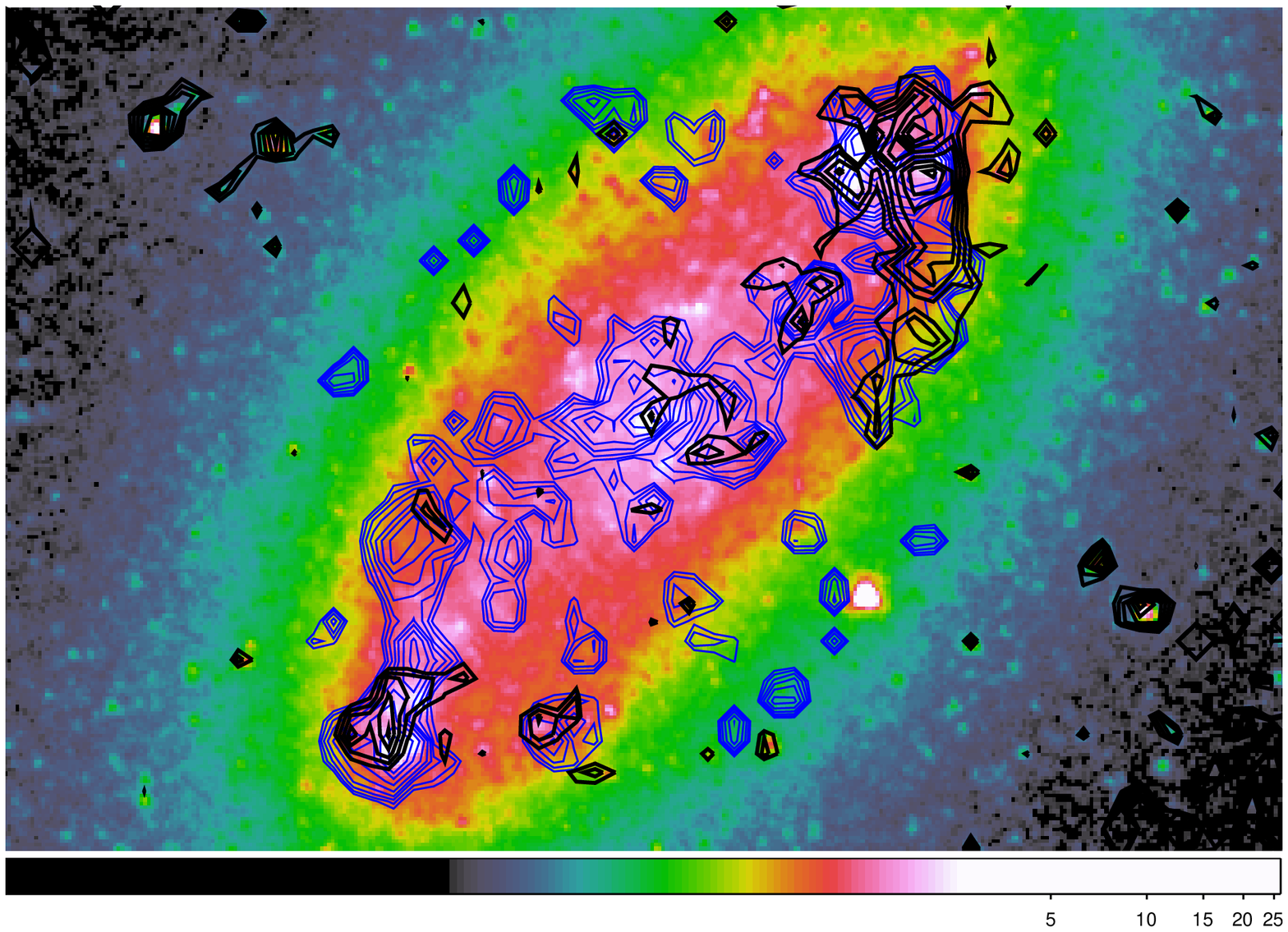}

\caption{\label{fig:image_1} Morphological Bar Evidence. Upper panel:  Color Map B-3.5 $\mu$, the curved dust lanes in the central region are characteristic of a weak/slow bar.  The upper right region suggest the geometry of spiral arms. Lower  panel shows the  3.6 $\mu$ image tracking the stellar component with few dust extinction, combined with dust map contours (thick black) and CO intensity contours (blue thin). The central region presents a box like structure resembling a weak stellar bar and is spatially coincident with the curved dust lanes. The contours for integrated intensity CO J=0 $-> $1, CO J=2 $-> $1, along each line of sight considering regions with $I_{CO} > 2\sigma$ (taken from Bolatto et al. in peparation and \citet{Leroy09}). Notice the change in position angle and the coincidence with boxy stellar structure as well with the dust lanes. }
\end{figure}

\begin{figure}[h!]
\epsscale{0.80}
\plotone{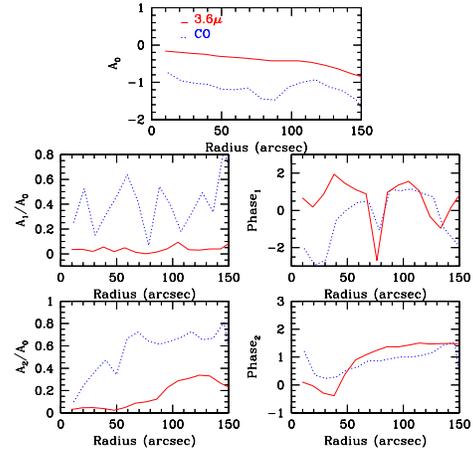} 
\caption{\label{fig:rmodes} Azimuthally averaged 1st and 2nd fourier modes. Top panel: Mean surface brightness profile in stellar component (solid) and molecular gas CO (dotted).
Medium left: A1 fourier mode (lopsidedness). Medium right: A1 phase profile in radians. Lower left: A2 fourier mode (bisymmetric distorsion) amplitude. Notice the A2 mode amplitude in CO is large while in 3.6 $\mu$}  is negligible, suggesting the bar gaseous richness. The A2 mode phase shows a big jump  close to 50 arcsec (0.5 kpc), coincident with the begining of spiral arms
\end{figure}

   Recently, it has been shown that bar instabilities are present in some dwarf LSB galaxies used to study the central dark matter distribution using their rotation curve (NGC 3109, NGC 6822 \citet{valenzuela07}).  The bar dynamical effects  are arguably enough to explain the discrepancy with cosmological predictions, at least when they act together with pressure support  \citep{valenzuela07}. For one dwarf galaxy (NGC 6822) the bar  has been clearly revealed in HI kinematics residuals after rotation substraction (see figure 2 in  \citet{Rhee09}, Rhee et al . in preparation) questioning if the rotation curve fully constrains the mass distribution for this galaxy. It is however necessary to analyze more dwarf galaxy cases cases before arguing that this is a general result.  On the contrary, NGC 2976 belongs to a sample of galaxies that has been argued to present a pure disk, with no bar, spiral arms or bulge,  the clear detection of  non-circular motions raised-up  the possibility of  being witnessing an elliptical disk triggered by the dark halo triaxiality \citep{Simon03, Simon05}.  Lately, Spekkens \& Sellwood (2007)  reported evidence favoring  a kinematic bisymmetric distortion over an inflow/outflow, leaving open the question if the kinematic perturber is a stellar bar or a prolate dark matter halo. More recently  \citet{Adams2976} found stellar kinematic evidence for non-circular motions possibly consistent with a disk inside a cuspy halo.   Although recent analysis of 2MASS images of NGC 2976 \citep{Menendez07}  marginaly suggested the presence of a stellar bar candidate, more recent theoretical studies conclude that based on the photometrically estimated  disk mass, the triaxiality must have been preserved after the disk formation \citep{KazantzidisTriax}, making appealing the triaxial halo presence scenario.  It is then important to quantitatively asses the bar presence/absence,  together with measurements of its strength. A possible bar may represent evidence of secular evolution in dwarf irregular galaxies, which importance is currently uncertain in evolution models.  In this paper we used techniques commonly used in giant high surface brightness galaxies, in order to confirm the bar presence and also with the aim to measure the strength of the perturbation with enough signal to noise.  The paper is structured as follows:   Section \ref{sec:obs} describes the observations used in this study.  Section \ref{sec:analysis}  describes the different diagnostics performed in NGC 2976  in order to detect and measure the bar/arm properties. Section \ref{sec:noncirc}  presents a discussion about the nature of non-circular motions and their relationship with a bar or non spherical halo.    Section \ref{sec:environment}  briefly discusses  the consequences for the detected bar and arms  for satellite galaxies transformation models in groups like tidal stirring. Finally   in section \ref{sec:concl} we present our conclusions.

 \section{Observations}                                                     
\label{sec:obs}     

\citet{Daigle2006} describes NGC 2976 as a peculiar dwarf ($M_B$ = 16.90) late-type (SAc pec) galaxy with a nearly linear rotation curve, no spiral 
arm visible and two strong HII regions located on each side of the galaxy. They report a Photometic/Kinematical P.A. and Inclinatios of 
323/323.5 and 63/70 respectively. Stil $\&$ Israel (2002) observed that in HI the rotation curve seems to flatten near the edge of the HI disc. 
According to Bronkalla, Notni $\&$ Mutter (1992), the outer parts of NGC 2976 have been undisturbed for a long time and are very old (5 Gy, 
probably up to 15 Gy).  This last conclusion seems to be at odds with the recent results in Walter et al. 2002 and Chynoweth et al. 2008 
who observed the M81/M82 group finding that M81, M82, NGC 3077, and NGC 2976 show the remnants of strong interactions as well as over 40 dwarf galaxies in their close neighborhood. Chynoweth et al. 2008 studied the HI emission of the group covering an area 
$3^{\deg} \times 3^{\deg}$ centered on M81 to 
include NGC 2976 and the extended emission associated the group finding an HI cloud located 27 kpc to the northeast of NGC 2976 and calculating 
a mass of 2.67 $\pm 0.65 \times 10^{7} M_{\odot}$ which may be contributing to the observed perturbation in NGC 2976. 

NGC 2976 is part of the Spitzer Infrared Nearby Galaxies Survey (SINGS) sample and our photometric analysis take advantage of the already 
available 3.6 $\mu$m IRAC images with a scale of 1.22 arcsec/pix. We combined these data with observations at visible wavelengths, mainly the 
$B_J$ band images of the photometric system by Gullixson et al. (1995) as observed by Frei et al. (1996) and scale of 1.35 arcsec/pix. 
We also use H$\alpha$ 
monochromatic maps from the Fabry–Perot of New Technology for the Observatoire du mont Megantic (FaNTOmM) with scale of 1.61 arcsec/pix and 
described in \citep{Daigle2006}.  After matching the images to the same resolution, we built a ($B-3.6$) color index 
map to get new insights on the structural properties of NGC 2976. For some of the analysis we tried to deproject the image assuming a circular 
disk and a disk thickness consistent with galaxies with similar rotation curve amplitude, we avoided the deprojection in some cases because 
the possible non-axisymmetric structure and inclination may complicate the traditional strategy \citep{Garcia-Gomez04}.

\section{Diagnostics for the Bar  Presence}                                                     
\label{sec:analysis}     

\subsection{Morphological Evidence: Dust Lanes, CO Intensity maps}

It is well established in  galaxy dynamics that a galaxy bar produces interstellar medium schocks wich can be traced by dust lanes \citep{Athanassoula92, Prendegast83}.
In order to seek for the presence of schocks we  constructed a B-3.6 $\mu$ color  image  presented in the upper panel of  figure \ref{fig:image_1}.   The dark structures are  the most dust obscured regions. Particularly notable are the pair of curved   dust lanes in the central region of NGC 2976, similar to the ones pesented by \citet{Athanassoula92} in the case of weak or slow bars. Near to the color map image upper right  spot there is a spiral arm like dust structure.  This finding supports the existence of a  disk bar and spiral arms, however it can still be argued that a random star formation pattern may originate the structures, therefore  we will study  other diagnostics like the kinematic response traced by H-$\alpha$ in the bar region in the following sections.  As a support to the bar interpretation we present at the  lower  panel of figure \ref{fig:image_1}, the CO integrated intensity contours in NGC 2976, taken from the HERACLES survey \citep{Leroy09} and the STING survey (Bollato et al in preparation) on top of the 3.6 $\mu$ image, and together with the color map contours. An elongated CO structure with different position angle compared with the average stellar component is outstanding, the structure twists on both sides developing narrow arcs similar to tightly wounded spiral arms. The spiral arms like features are  coincident with the dust map structures. The two symmetric bright spots suggest the beginning of spiral arms.

\begin{figure}[h!]
\epsscale{2.2}
\plottwo{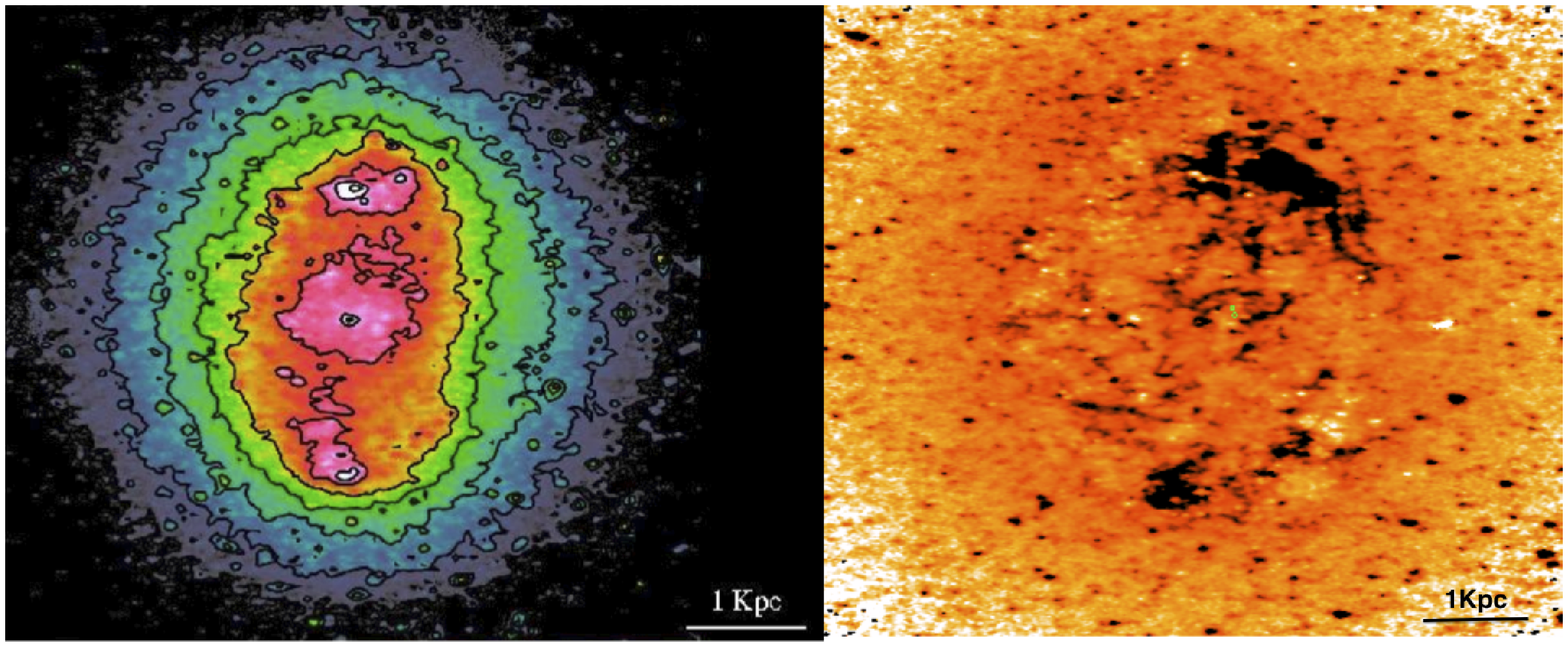} {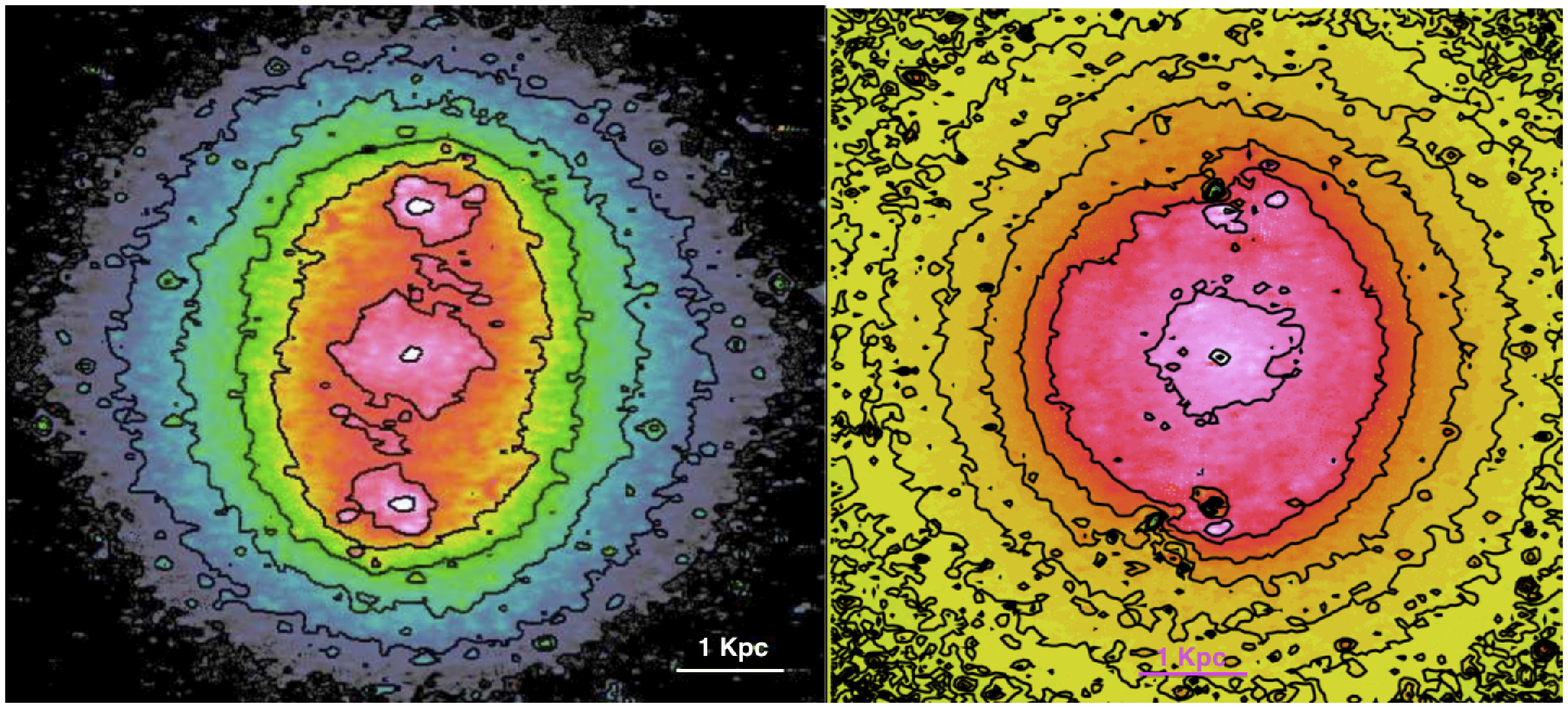} 
\caption{\label{fig:multi_panel} Even and odd  structure. The figure shows 4 images. Upper left panel presents the original Spitzer image in 3.6 microns, deprojected under the assumption of a circular external disk. Upper right hand side shows a deprojected B-3.6 color map. The lower left  presents an image constructed only with the even powers of a 2-dimensional fourier decomposition. Notice the spiral/oval structure. The Lower right hand image, shows am image constructed with only odd powers.  The circular yellow contours suggest the the galaxy disk is not lopsided in the external region. }
\end{figure}

 \begin{figure}[h!] 
\epsscale{1.00}
\plottwo{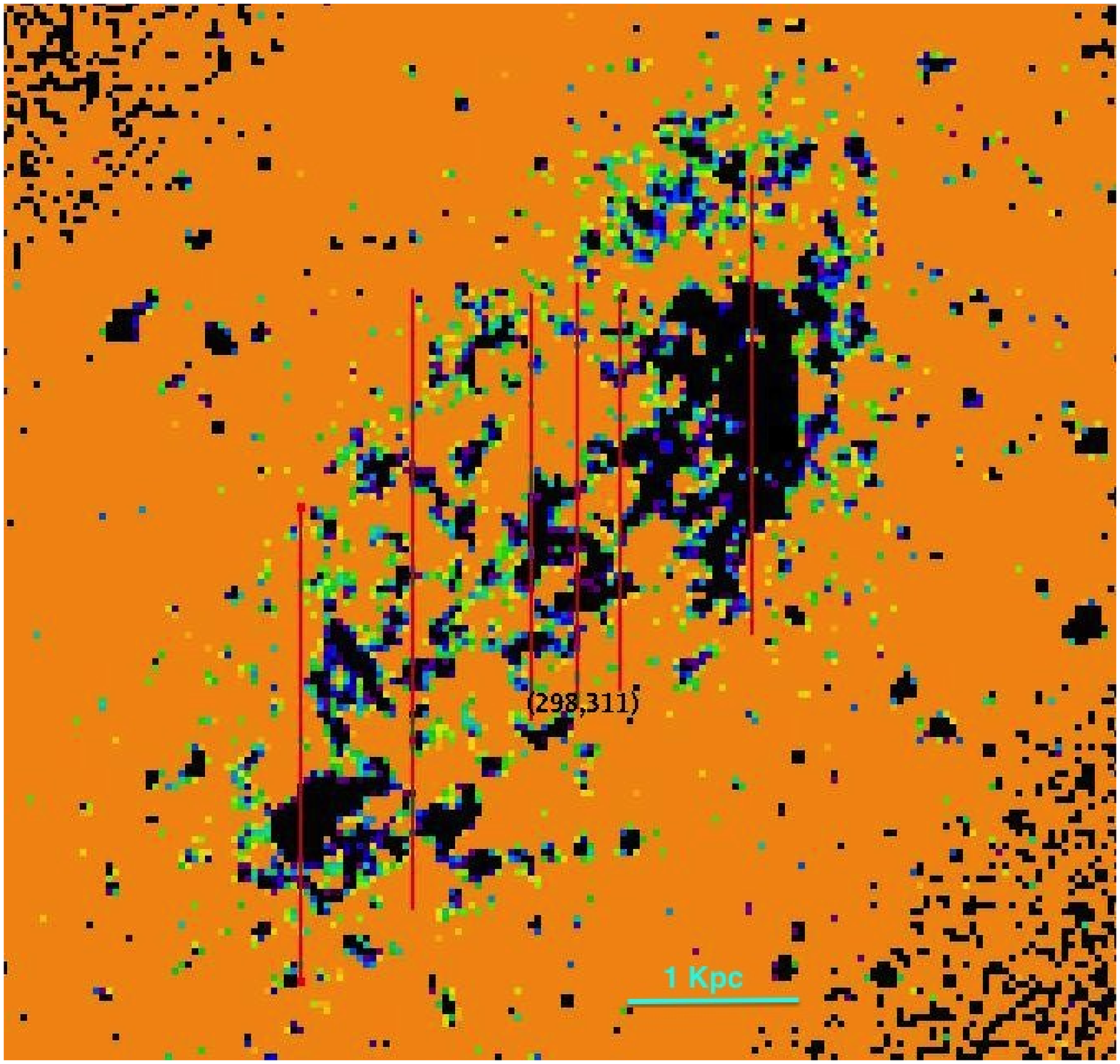}{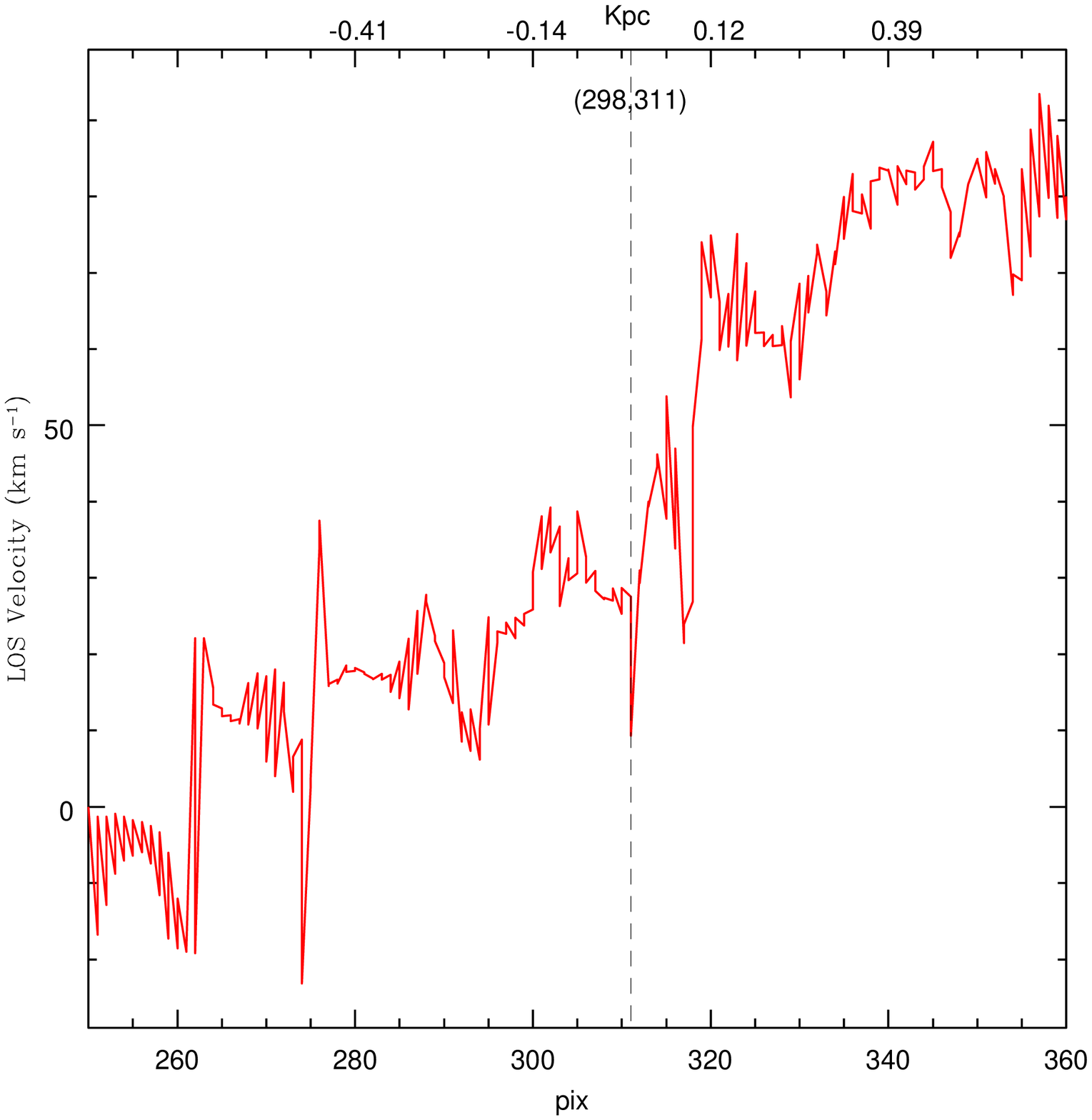}
\caption{\label{fig:shocks} Dynamical Effect of the Bar. Left panel presents the color map showing  dust lanes across the whole galaxy. The vertical lines show cuts along the velocity field crossing the dust lanes. Specifically the vertical lane labeld 298, 311 crosses the curved dust lane at this pixel coordinates. The right panel  shows  how the velocity profile presents a jump of about 30km/s  going from pixel 330 to 311, in other words crossing the dust lane.  }
\end{figure}

\subsection{Radial Fourier modes}   
A bisymmetric structure like a galactic bar or a spiral arm pair  can be well represented by even fourier components. A first diagnostic  to the presence of such structures is the amplitude of the second fourier mode and the behaviour of its phase.
Figure \ref{fig:rmodes} shows the first two fourier  modes for the average stellar component and the CO gas emission (solid and dashed line respectively). The first fourier  mode shows a very small amplitude in the stellar component (less than 0.05), while for the CO tracer  the behaviour is noisier but with a larger amplitude. The phase is consistent for both tracers.  It is not straight forward to disentangle if this amplitude is the result of the clumpy distribution of molecular gas or a truly global lopsidedness.   The second fourier mode barely has a non zero amplitude inside 70 arcsecs in 3.6 $\mu$ where the isophotes show a boxy shape,  however its amplitude in CO indicates that  20$\%$ of the surface brightness is in the m=2 mode, consistent to the more elongated CO distribution.  The corresponding phase is  consistent in both tracers, showing a big jump at 50-70 arcsecs,  suggesting the transition from the bar to the spiral arms.   Azimuthal  average and  projection effects may have a non-trivial influence on one dimensional fourier analysis, not to mention the flocculent structure, therefore we will apply other diagnostics.

\subsection{Two Dimensional Fourier Analysis}   
In order to keep azimuthal information we performed a 2-Dimensional
fourier analysis of the 3.6 $\mu$ image. We splitted the image in two: one image built with all the even modes and another  one built with the odd modes.   If there is a bar or bisymmetric arms  it must be shown up in the even modes image. Figure \ref{fig:multi_panel}   shows at the top the deprojected 3.6 $\mu$ image next to a de-projection of the color map and at the bottom both even and odd images (left and right respectively). The even image clearly shows a boxy structure at the position of the CO bar and the beginning of bi-symmetric spiral arms. Although encouraging  complications associated with the de-projection  have to be carefully handled \citep{Garcia-Gomez04}, it is however remarkable that bar and spiral arm structures are so well defined despite the radial Fourier mode in the stellar component is not large, supporting the interpretation that both bar and arms are  gas rich but they have a stellar counterpart.

\subsection{Kinematic Signatures of Bar Induced Shocks}
Shocks across dust lanes are readily identified as steep velocity gradients \citep{Athanassoula92, Weiner2001}.  We used the H-$\alpha$  velocity field kindly previously  discussed by \citet{Daigle2006}. The data points in the right panel of figures \ref{fig:shocks}  show the observed velocities along a pseudo-slit crossing the velocity field placed with an orientation perpendicular to the curved dustlanes as shown in its left panel. The pseudo-slit reveals a projected velocity gradient of almost 30 km sÐ1 is coincident with the prominent dust lane. A similar configuration and signature is discussed in  \citep{Zanmar2008}. We conclude that regardless of projection effects, the gas-rich bar has a dynamical effect on NGC2976 internal kinematics.

\subsection{Two-Dimensional Force Map}
\begin{figure}[h!]
\epsscale{.90}
\plotone{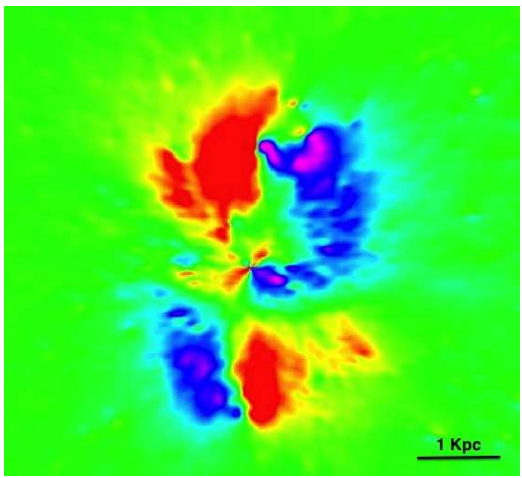}
\caption{\label{fig:torque}.Torque Map based on the 3.6 $\mu$ image and solving the Poisson equation. Red (Blue) colors correspond to negative (positive sign). The alternating sign is the distinctive  signature of a bar/arm quadrupole.   }
\end{figure}

Once we are confident about the presence of a bar in the stellar component of NGC 2976, it is natural to ask for the bisymmetric perturbation strength. \citet{Buta2001}  introduced the ratio of  tangential to radial force $Q_{b}$ parameter  as a diagnostic of the bar/arm strength.  After solving the poisson equation on the deprojected stellar component image,  the potential is diferentiated in order to calculate radial and tangential forces  point by point.  Afterwards we built a two dimensional map of the  ratio as is shown is figure \ref{fig:torque}. The characteristic  bar/arms  signature is the alternating sign of this ratio as we move through quadrants, verifying the bar/arms interpretation in our previous diagnostics.  The central region shows a slightly asymmetric central pattern pattern and the large scale diagram ilustrate the spiral arms force as is discussed by  \citet{Block2004}   The ratio of the maximum tangential force to the average radial axisymmetric amplitude has been shown as an amplitude measurement   \citep{Combes1981}, for NGC 2976 the value in the bar region is around 0.4, therefore regardless of its nature the bi-symmetric perturbation is a strong one. At the arms region the tangential perturbation is even larger reaching values of 80$\%$ of the radial force.

 \section{Dark Matter Distribution}    
 \label{sec:halo}     

NGC 2976 dark mater distribution has been discussed by several studies,  most of them favored  a cored halo distribution based on interpretation of the rotation curve \citep{Simon03, Spekkens07, DBlok2008}.  However, recently  \citet{Adams2976} concluded that stellar kinematics and anisotropic Jeans modeling combined with information of its  star formation history favored a cuspy dark matter distribution. A critical difference between these studies is the interpretation of non-circular motions. These non-circular motions are detected both in gas and stellar component. In the next sections we will  discuss their possible nature.

\subsection{Non Circular Motions:  Triaxial Halo/ Bar-Arms?} 
\label{sec:noncirc}     
NGC 2976  has been proposed as a system that provides  kinematic  evidence for a  triaxial halo. This is important because the  existence of non-spherical dark matter halos  is an ubiquitous prediction of hierarchical  structure formation model. Galaxy disks are expected to react to the triaxial halo potential become elliptical and even develop instabilities triggered by the halo perturbation,  the halo triaxiality is also modified by the process, making non-trivial the predicitons about the central density structure \citep{Bailin07}.  However, a positive detection of halo triaxiality in disk galaxies has been hampered partly because of the degeneracy between true disk  ellipticity and projection effects. A promising avenue to verify this prediction is a kinematic detection of disk ellipticity in the absence of bars and spiral arms. \citet{Simon03}  pointed out that NGC2976  shows considerable non-circular motions but the galaxy lacks of an obvious non-axisymetrics internal baryonic perturber, raising the possibility that we are witnessing an elliptical disc triggered by a  triaxial halo. Spekkens \& Sellwood 2003 showed that a bisymmetric model is  preferred by the data over a radial flow. Furthermore, recently Kazantzidis et al 2010 showed that  constraints on the disk mass suggest that  adiabatic disk formation would have not been able to erase halo triaxiality.  Although our results  do not question the theoretical calculations,  we positively detect a bar and spiral arms.  It is fair to mention that the bar and arms system is unusual, the bar is clearly shown in CO intensity map, but in HI  the bar is rather uncertain but arms are detectable. In 3.6 $\mu$ the bar hint is only a boxy like structure, and the arms are noticeable mainly through the symmetric bright spots. The situation in normal galaxies is the opposite, a clear stellar bar is observed and some gaseous response to the dynamical perturbation hinting the bar.  If NGC 2976  gaseous disks is dominant over  stars  is possible to explain its morphology. However, the reported stellar mass is $5 \times 10^{8}$ M$_\odot$  and the gaseous mass at most 30\%  of stars  ( $1.5 \times 10^{8}$ M$_\odot$)\citep{delpopolo2012}. Although there is evidence of gas tidal stripping therefore the galaxy may have been even richer in gas, there is not  obvious reason of why the bar should be more visible  in gas. \citet{W2010} and others have discussed the presence of a stellar spheroid or halo of intermedia age stars, this structure may be observed in projection as a boxy bulge.  One possible alternative interpretation is  that  indeed we are witnessing the gas/stars response to an elongated dark matter halo, in such a case stars reacted to the triaxial halo orbital structure developing the boxy bulge. Indeed, disk  non-axisymmetric structures and triaxiality are not necessarily  mutually exclusive. It is well known that an elongated halo is able to trigger spiral arms \citep{Bekki2002}   Some recent studies  simulated the interaction of bars and disks inside triaxial live halos, concluding that disk bar formation is possible but it can  also erase  halo triaxiality in the central region (Berentzen et al 2005, Athanassoula et al 2010).  However the models were tuned to high surface brightness galaxies unlike NGC 2976, therefore because of its low  baryonic fraction, we can not rule out the possibility that considerable triaxiality  has survived in the central region of NGC 2976 regardless of the bar/arms presence.  We conclude that non-circular motions in the central 80 arcsecs are likely triggered by the detected  bar/arms, and is at least contrived to compare the importance of bar/arms against halo triaixiality inside 50 arcsecs.  Nevertheless,  triaxiality may survive at the galaxy outskirts. In order to test this interpretation we used controlled SPH simulations of gaseous disks in triaxial halo potentials. The simulations will be described in detail somewhere else (Pichardo et al in preparation), however here we present only the basic information. The halo has a logarithmic potential with a very small core radius, mimicking a cusp-like  halo model.  A Kuzmin gaseous disk with an isothermal equation of state is initially set in equilibrium  inside an spherical halo,  later during 20 dynamical times at twice the initial disk  radius (adiabatically),  the halo model axis ratios are gradually modified, recently a similar study was described in the literature \citep{Khoperskov2012}.   The disk responds developing ellipticity, and spiral arms, and the gas inflow triggers a bar like central structure which realism is uncertain because the idealized simulation set up.  An interesting finding is that for disk plane ellipticity of  0.3-0.4 or larger  the gas at turn around piles up and in  the more eccentric cases develops two symmetric shocks. NGC 2976 shows two symmetric bright spots in the  stellar photometry and H-alpha, suggesting that either the spiral arms are stronger than expected  or the disk ellipticity  at that radius  is near  0.3-0.4  or greater.

 \begin{figure} [h!]
\epsscale{.45}
\plotone{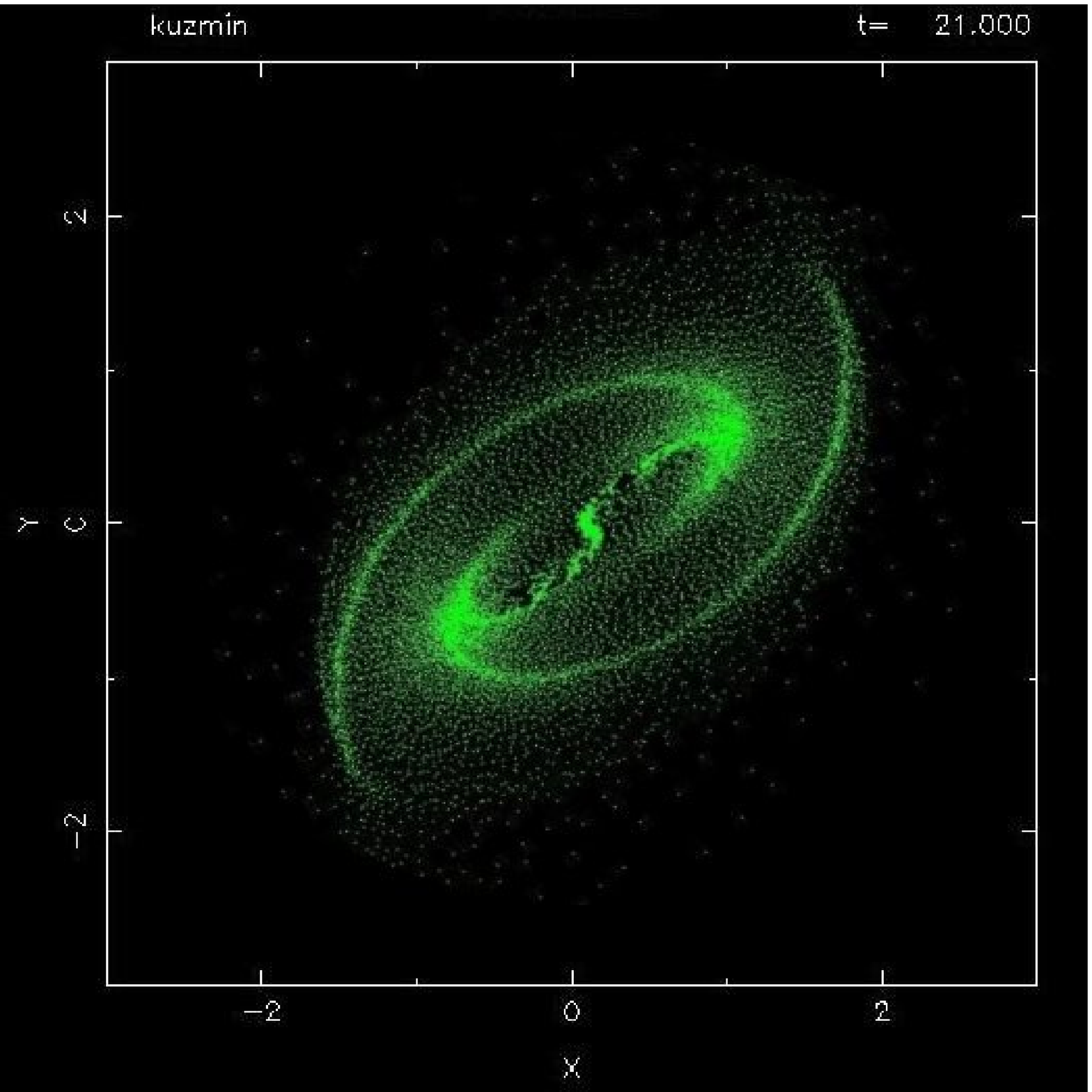}
\epsscale{.45}
\plotone{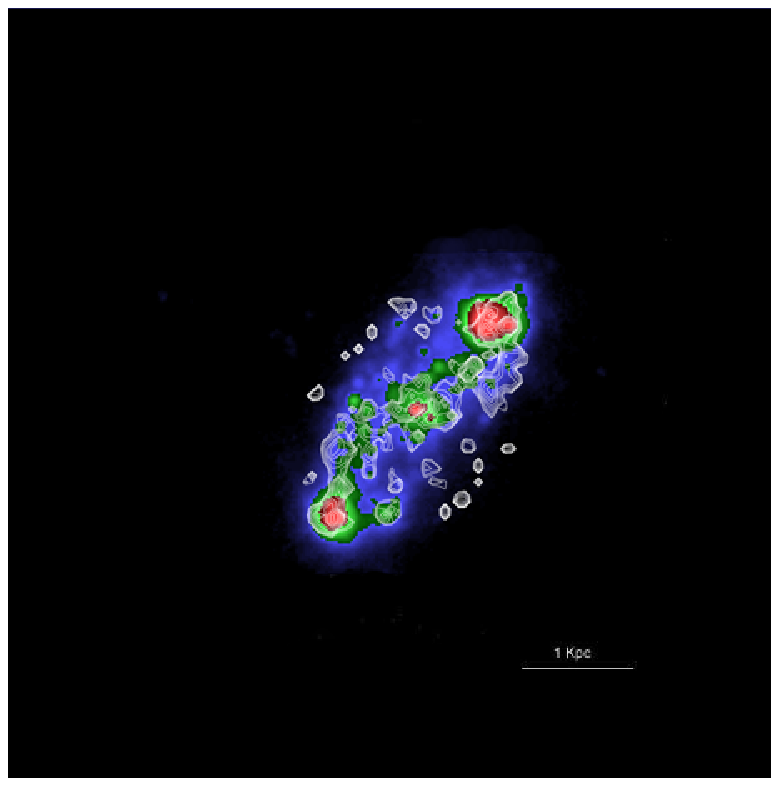}
\caption{Symmetric Bright Spots. First panel: SPH gas simulation of  a Kuzmin disk  inside a triaxial Dark Matter halo. The disk was introduced adiabatically, after some dynamical times,  spiral arms are developed and gas at turning points piles up along the line of nodes creating an over-density and either a shock for ellipticities larger than 0.5. Notice the symmetric overdense regions where presumably star formation may happen. Second panel: 24 $\mu$  Spitzer/MIPS image tracking dust warmed by star formation. Noticed the two symmetric bright star formation regions. They could be evidence of gas turn around triggered by a  prolate dark matter halo as in the simulation,  or  alternative by strong bar/spiral arms.}
\end{figure}

 \subsection{Environment Triggered  Evolution  of  Satellite Dwarf Galaxies}    
 \label{sec:environment}     
 Once we have confirmed the presence of  a gas rich non-axisymetric structure in the disk of NGC 2976, the most natural explanation is that the  tidal interaction with the environment triggered a mass redistribution, making the galaxy unstable regardless of the low baryonic mass fraction.
 Specifically, N2976 environment  is the M81 group,  an extended HI tidal tail suggests that  the galaxy has indeed suffered an interaction with the M81-M82 system.     
The whole  M81 group mass is  close to $10^{12} \Msun$ (Karachentsev et al. 2006) and the projected distance  up to NGC 2976 is close to 190 kpc, therefore a crude estimation for the
crossing time is 1 gigayear. \citet{Chynoweth2008} estimates 0.2-0.3 gyrs for the age of the neutral gas bridge connecting M81-M82 which based on the bridge projected extension, is younger than the one connecting M81 and NGC 2976, therefore a 1 gigayear estimate for their interaction age is    conservative and also in agreement with \citep{W2010}  In comparison, at the last measured optical rotation curve point (2 kpc) the NGC 2976 rotation amplitude is 80 km/s, therefore the  rotation period is approximately 0.024 gigayears, suggesting that enough time has spent since the interaction  in such a way NGC 2976 disk should be  relaxed by now.  We conclude that  disk self-gravity is important in NGC 2976 and as a consequence the bar and spiral arms are  unlikely transient structures. Recently a low luminosity AGN has been revealed in NGC 2976 through  X-Ray and NIR observations \citep{Grier2011}, this probably reveals the presence of gas inflows triggered by the non-axisymmetric structure and also of energy injection into the central galaxy ISM , both processes may have effect on the dark matter halo structure.

 \subsection{Generality}    
 \label{sec:generality}     
 
A critical question to ask is: How common is NGC2976 between  satellite dwarf galaxies?
The  are two ways to address such an inquire: through statistical analysis of satellite galaxy population
or using theoretical predictions. There are not many systematic studies searching for bars in satellite galaxies around normal galaxies.   Statistical analysis of more massive galaxy groups show that AGN«s  potentially triggered by secular evolution are  more common in satellites than in central galaxies \citep{Allevato2012},  however there are not many similar studies  in less massive groups like M81. It is natural to ask if the whole evolution including disk instability and nuclear activity  is triggered by the environment and if a similar situation may be common to other satellite dwarf galaxies located in low density groups like M81 or the Local Group.  Recently \citep{Grier2011} found AGN candidates in a sample of THINGS galaxies where NGC 2976 is included, however more work is required.     From the theoretical    side, the scenario  is a low red-shift incarnation of the so called  tidal stirring one (Mayer et al 2001).  \citet{KazantzidisTidal2010} recently characterized the evolutionary tracks of satellite dwarf galaxies finding that bar instabilities developed before a global mass redistribution happened, the evolutionary stage seems quite insensitive to orbit details and more dependent of the cumulative tidal force\citep{Lokas2010}, then is  more sensitive to the satellite accretion time.  However, more recent calculations including SN feedback of gas rich dwarf galaxies, seem to show that no bar instabilities developed in galaxies similar to NGC 2976, likely because of gas inflow may inhibit bar grow \citep{Mayer2011}.  Finding a bar and spiral arms in NGC2976 is a good constraint to this model. Based on NGC2976 we can especulate that the energy injection triggered by the low luminosity AGN/nuclear starburst  may constraint the gas inflow allowing the bar growth.  Further insight into this matter may be obtained analyzing more satellite galaxies in M81 and other nearby low density groups (Hernandez-Toledo, Valenzuela et al in preparation), as a matter of fact some hints suggesting environmental transformation of satellite galaxies in low mass groups may be already reported in M101 group \citep{Mihos2012}. Another consequence of this scenario is that the transformation process triggered by environment may bias comparison against theoretical predictions for isolated galaxies, including expected mass-Vmax-concentration relationships, triaxiality, star formation rate \citep{Colin2010, Avila2011, Aldo2012}  and how these properties correlate to the baryonic structure of the galaxies residing inside dark matter subhalos.


\section{Conclusions}      
\label{sec:concl}     

Four different indicators: CO intensity map morphology, curved central dust lanes, fourier m=2 mode surface brightness decomposition in 1-D and 2-D and H-$\alpha$ kinematic jumps across the central  dust lanes, all suggest that the NGC 2976 disk shows a non-axisymetric structure:  a  gas rich central bar structure and gaseous spiral arms at large scales.

The gas rich bar/arms is expected to produce non-circular motions biasing constraints on the mass distribution estimation in agreement with recent kinematic analysis of NGC 2976 \citep{Adams2976}, similar structures may bias comparison of this galaxy versus cosmological predictions for isolated galaxies \citep{Aldo2012}.

The bar and  arms are the most natural triggers of the observed non-circular motions, mass redistribution and nuclear activity(AGN/nuclear starburst, \citet{Grier2011} ),  however a detailed modeling of the galaxy and its environment is required in order to decide between environment induced bar/arms and triaxial dark matter halo induced elliptical disk and arms.  

If the bar/arms are not as strong as  CO observations suggest, the two symmetric spots 
visible in H-$\alpha$, and infrared, are possibly indicating an elliptical disk with ellipticity greater than 0.3.

The  bar existence is consistent with tidal stirring predictions, the recently reported low luminosity AGN/nuclear starburst may be connected with the bar and may be a common stage on satellite galaxies evolution,  with possible influence in the star formation history.  

  Overall, NGC 2976 considered inside a sample of simple,  isolated, axisymmetric dwarf galaxies, is revealed as a complex galaxy affected by its environment and showing peculiar spiral arms and bar. Secular induced mass redistribution is likely important and connected with the reported AGN/nuclear starburst activity,  the same conclusions likely apply to the evolution of many  other  dwarf satellite galaxies.

\section{Acknowledgments} 
We thank P. Amram for kindly making available the fabry-perot data used in this paper. H. Segura is acknowledged for help with graphic material. We thank valuable conversations with: G. Rhee, J. Simon, J, Navarro, P. Teuben,  A. Bosma, K. Spekkens.
Support by the UNAM PAPIIT IN118108 grant is acknowledged. OV acknowledges support from grant PAPIIT/UNAM: IN112313. MCD was supported by the Marie Curie Training Network ELIXIR under the contract PITN-GA-2008-214227 from European Commission.

\end{document}